# Comparison About EEG Signals Processing in BCI Applications

Giulia Cisotto, Silvano Pupolin
Dept. of Information Engineering
University of Padua
Padua, Italy
{cisottog,pupolin}@dei.unipd.it

Francesco Piccione
Dept. Of Neurophysiology
I.R.C.C.S. S. Camillo Hospital Foundation
Venice, Italy
francesco.piccione@ospedalesancamillo.net

*Abstract*—In the context of a Brain Computer Interface platform implemented for the arm rehabilitation of mildly impaired stroke patients, two methods of EEG signals processing are compared in terms of (i) their identification performance rate and (ii) their computational complexity with the overall goal to select the most efficient and feasible real-time procedure. An effective signal processing is, indeed, one of the most critical issue for such kind of technology which aims to establish a real-time communication between the subject's brain and a machine, i.e. a computer, a robotic arm or another device, that should implement his/her intention to move in place of his/her impaired arm.

## I. Background

Originally implemented for the augmentative communication of completely-locked-in patients that did not retain any capacity to communicate with the external world, Brain-Computer Interfaces (BCIs) or Brain-Machine Interfaces (BMIs) have been recently thought back also as auxiliary tools to accomplish movements that such paralyzed people could not perform anymore [1,2,3,4,5]. The basic underlying principle of a BCI or a BMI is to establish a mutual dependence between the subject's brain and a computer in order to make (i) the subject trained in using the machine to his/her needs and (ii) the machine to recognize the specific subject to help him/her in communicating, moving a robotic arm or driving a wheelchair. Control signals exploited in such systems to drive the BCI following the patient's intentions could be acquired in many ways. However, thanks to its high temporal resolution of the order of milliseconds, its cheapness and its non-invasiveness, the electroencephalogram (EEG) has been nowadays widely recognized as the most promising acquisition system to integrate in at-home and portable future BCI systems that every patient could own. Spontaneous or induced modulation of particular EEG components around 10 Hz (μ) and 20 Hz (β) in the sensorimotor scalp areas, namely *sensorimotor rhythms* (*SMR*), were observed to be movement-related: during rest periods (*idling state*) indeed, they usually showed high amplitudes, while they became suddenly suppressed (desynchronized) when an action was going to be performed, or when it was imagined, observed or actually performed. This phenomenon was defined as *event-related (de)synchronization (ERD/ERS)* [6] and it is widely used in BCI applications devoted to restore movement [7]. When a subject accomplishes to a motor task driven by its cerebral activity, indeed, a strategy to promote the correct neural pattern in correspondence to the movement has to be exploited: in this case, the *operant-conditioning* is used [8]. Thanks to this paradigm, when desynchronization of the SMR is observed just before the motor task, a motor output can be performed by a robot in place of the paralyzed arm of the subject. EEG signals processing can then be easily identified as one of the most important elements in the implementation of an effective system. In fact, a proper algorithm should be implemented to precisely and reliably detect the desynchronization of SMR in real-time before the movement for each subject, even across the inter-subjects and inter-trials variability. Besides, a preliminary detection and suppression of artefactual, i.e. non-movement related, components in the incoming EEG signals has to be integrated in the online procedure as well. In the paper two signal processing methods, the one proposed by Pfurtscheller and colleagues [6,9,10] and the other by authors will be compared in terms of (i) their identification performance rate and (ii) their computational complexity with the overall goal to select the most efficient and feasible real-time procedure with the perspectives highlighted before.

## II. Materials

As mentioned above, two signal processing methods are going to be presented and compared. They were tested on a dataset acquired by means of a specific platform [11] made by a BCI and a haptic device that aimed to help the patient in performing a standard reaching task on a plane in one out of four cardinal point at any time whenever they modulated their SMR in a satisfactory way (in terms of amplitude and time). In the following the platform will be briefly presented along with some details about the timing of the protocol that are necessary to the sake of clarity for the subsequent comparison.

### A. BCI-Phantom platform

As it can be observed from Fig. 1, the system was formed by the four blocks of a typical BCI: a signal acquisition unit, a signal processing module, a feedback control unit and a force feedback provider, the Phantom device.

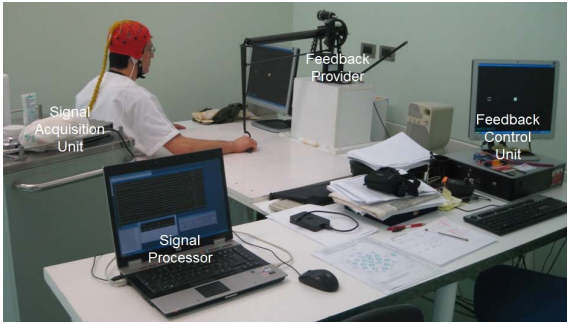

Fig. 1. The BCI-Phantom system.

In particular, 16 EEG channels were acquired by means of a gTEC amplifier with the electrodes selected from the sensorimotor areas of the subject's scalp, as illustrated in Fig. 2.

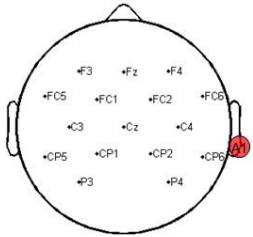

Fig. 2. The EEG montage.

Then, each signal was digital-to-analog converted with a 512 Hz of sampling rate and a 24 bit of quantization. A band pass filter in the band (0.1, 60) Hz was also implemented in the hardware along with a notch filter in 50 Hz to avoid the mains interference. BCI2000, a world-wide used software, was exploited to acquire, process and extract few relevant features in real-time from the incoming EEG signals. Those quantities were finally sent to the third block of the system where the feedback control unit transformed them into suitable feedback outputs, i.e. a force level between 0 and 6 Newton. Accordingly with the operant-conditioning strategy cited in the introductory part, a real-time feedback is needed during the operations in order to strengthen the concurrence of the voluntary desynchronization of the SMR modulated by the patient and the correct reaching movement performed by the subject with the help of the robot. Specifically, the Phantom device was activated to help the patient in completing the movement whenever he/she was desynchronizing his/her sensorimotor activity, i.e. showing the intention to move.

*B. The dataset*

The dataset was collected by a series of measures through the all 24 BCI *sessions* of the experiment, covering a period of approximately three weeks. Three *runs* made by a set of 80 *trials* per run were recorded with some of the trials excluded from the analysis because of the occurrence of artefactual events during their acquisition. A stroke patient suffering from a left-side motor impairment was involved in the experiment and was required to perform it both with his left impaired arm and the right healthy one. As a comparison, a healthy subject was also recruited in the experiment to verify the robustness of the system during the first steps of the implementation of the platform. For the sake of completeness, further details about the system and the patient characteristics can be found in [11]. Nevertheless, it is worth to mention the timing of a single trial of movement because of its importance in the establishment of the comparison between the two different signals processing methods. Specifically, the subject sitting in front of a screen displaying the graphical interface of Fig. 3 was required to hold the end-effector of the Phantom device as captured by Fig. 1 and, starting from the central position marked by the green circle in Fig. 3 moving to the target, randomly selected by the procedure and shown as a white square in one out of the four cardinal points of the screen.

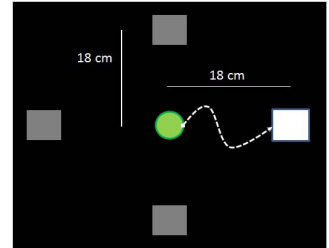

Fig. 3. The graphical interface of the reaching task.

Particularly, the timing of the task imposed the patient to rest for 500 ms at the beginning of each trial (*pre trigger time*); after a first cue sound given when the target to reach appeared, he had to wait for 1500 ms in the starting position (*post trigger time*) until a second cue sound was given. After that, he was allowed to move towards the target: he took 500 ms approximately to react (*reaction time*) and the subsequent movement (*movement time*) had to be performed within the range (500, 740) ms in order to be considered correctly accomplished. A *recovery phase* was accounted for the patient to return back from the hit target to the starting position. Such a timing is schematically reported in Fig. 4.

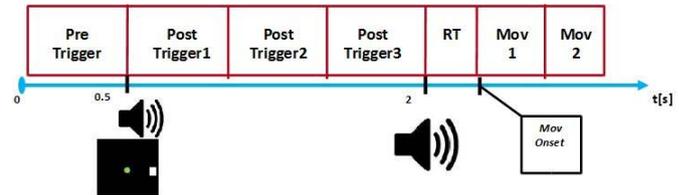

Fig. 4. The timing of the reaching task.

Consequently, it clearly results that the most crucial element in the protocol in order to realize the operant-conditioning strategy is to extract the most characteristics features of the movement from the incoming EEG signals and to provide in real-time (before the reaction time beginning) a suitable force feedback.

III. METHODS

Since a comprehensive study of the desynchronization characteristics of the EEG signals of the specific dataset available by the authors had not been performed yet, two methods were compared in order (i) to describe the individual patterns related to the movement and (ii) to highlight the most suitable method to be implemented in real-time in terms of computational complexity and identification performance.

A common preliminary stage for the electrode-polarization artifacts removal was operated. The algorithm, implemented by authors, identified and removed large artifacts like the one of Fig. 5 operating in three steps: (i) the derivation of the signal, (ii) the identification of the artifact based on the detection of some consecutive samples with abnormally large amplitude values of the derivative and (iii) the non-linear suppression of the short signal segment between two consecutive zero-crossing points around the spike of the signal derivative. The subsequent narrow-band filtering actions was then no longer affected by the artifact effects. Moreover, in this way during the short time of about 1 second when the signal was suppressed, the robotic feedback was suspended instead of giving a completely random stimulus to the patient due to the artifact effects on the EEG features extraction. After this step, one out of the following methods could be applied.

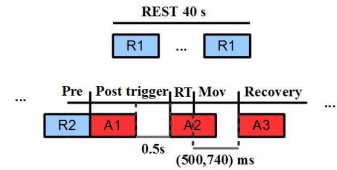

Fig. 6. Timing of the analysis with the standard method.

Then, for what the first step of the method concerns, there are several ways to determine the subject-specific frequency band, as reported in [9]: (i) the detection of the most reactive frequency band based on the comparison of two short-term power spectra, (ii) a continuous wavelet transform based method and (iii) the definition of frequency bands relative to the spectral peak frequency. In this context, the first method was exploited. Particularly, the difference between the mean logarithmic power spectrum of each reference period and the analogous spectrum computed during an active period was estimated along with the 95% confidence interval. 2 Hz frequency bands were identified with each of them selected as the interval in which a positive-valued difference exceeded the confidence level. Afterward, each trial was band-passed in the correspondent individual frequency band, each sample of it was squared to obtain power samples and, finally, an average trial made by power samples was computed. Then, the mean power into each period $R_1$, $R_2$, $A_1$, $A_2$ and $A_3$ was calculated and the ERD/ERS between each pair of a reference and an active period was estimated through the formula: (A–R)/R*100. To investigate the ERD/ERS patterns, recalling the literature about the motor imagery tasks, C3 was taken into account for right-side movements of the hand or the arm, C4 for left-side movements of the same limb segments and Cz for the movements of foot and tongue. Therefore, in this analysis C3 and C4 were mainly considered. Cz was also accounted for completeness.

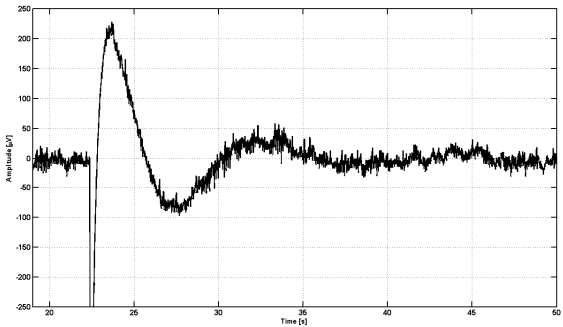

Fig. 5. An example of electrode-polarization artifact.

### A. The standard method

The first method was proposed in the 70s and later on revised by Pfurtscheller and colleagues [6,9,10] and represents the almost standard way to compute the ERD/ERS phenomena of the SMR. It is mainly constituted by two steps:

I. the selection of an individual frequency band;

II. the ERD/ERS computation in the frequency band previously identified.

It has to be recalled that the quantification of the ERD/ERS rely on the measurement of the changes in the power of the SMR bands just before a movement. This is accomplished by computing the power of such rhythms in both a rest period (R) – when the subject is relaxing – and in an active one (A) – when the subject is planning, performing, observing or imaging a movement. In order to adapt the Pfurtscheller's method to the data obtained from the experiment described before, two one-second R periods and three one-second A intervals were selected. This led to a more comprehensive comparison between the two methods. In particular, $R_1$ was selected during the initial long-lasting rest, while R2 represented the set of the inter-trials periods previously named as pre-trigger times. As the active periods as regard, $A_1$ accounted for the planning phase, $A_2$ for the actual movement and $A_3$ for the recovery phase. More specifically, the structure in Fig. 6 reports this timing.

### B. The novel method

On the other hand, the method proposed by authors was made by 5 steps:

I. 33 differential signals were calculated as the difference between pairs of near channels;

II. 2 Hz wide frequency bands were chosen to cover the range (5.5, 16.5) Hz with subsequent bands overlapping by 1 Hz;

III. 500 ms intervals delimited by triggers of the experiment (refer to Fig. 7) were selected to subsequently compute the ERD/ERS;

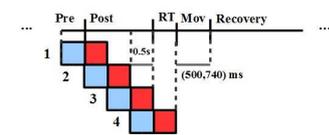

Fig. 7. Timing of the analysis with the author's method.

IV. the ratio between the energy in each interval before the onset of the movement and the previous interval in

each signal and each band was computed. Four definitions of ERD were then performed: post-trigger 1/pre-trigger, post-trigger 2/post-trigger 1, post-trigger 3/post-trigger 2 and reaction time/post-trigger 3;

V. an ERD identification was accounted for a specific ratio and a particular band if the mean ERD value among signals of the same group was less than 60%, that is an energy decrease of 40% at least occurred from an interval to the subsequent one, showing a significant ERD.

## IV. RESULTS

In the following, results from the application of the two methods to the available experimental data will be presented. Data was collected during the reaching movement of the left-paretic arm. In particular, for each pair of reference and active periods the individual frequency, the percentage of identification of an ERD stronger than 40% and the mean ERD value along all the trials will be reported for the first methods. C3, Cz and C4 were distinguished among the results. Results were computed over a set made by 2191 trials. On the other hand, the analogous values of identification and ERD entity will be shown for the second algorithm. In this case, the mean values within each group of signals (left-side, inter-hemispheres, right-side) were evaluated and then compared with results from C3, Cz and C4 of the first method, respectively. In this case, results were computed over a set made by 1912 trials.

TABLE I. INDIVIDUAL FREQUENCY FOR EACH PAIR OF INTERVALS R, A (STANDARD METHOD).

|  | C3 |  | Cz |  | C4 |  |
|---|---|---|---|---|---|---|
|  | *Mean (Hz)* | *Std (Hz)* | *Mean (Hz)* | *Std (Hz)* | *Mean (Hz)* | *Std (Hz)* |
| $R_1A_1$ | 11,84 | 1,26 | 11,26 | 1,24 | 10,95 | 1,13 |
| $R_1A_2$ | 8,21 | 2,37 | 8,00 | 2,05 | 7,96 | 2,35 |
| $R_1A_3$ | 10,93 | 1,73 | 10,67 | 2,11 | 10,27 | 1,33 |
| $R_2A_1$ | 11,25 | 2,30 | 12,00 | 2,00 | 11,87 | 1,85 |
| $R_2A_2$ | 6,87 | 2,10 | 7,04 | 1,52 | 6,61 | 1,08 |
| $R_2A_3$ | 10,20 | 1,75 | 9,91 | 2,02 | 11,27 | 2,05 |

TABLE II. PERCENTAGE OF IDENTIFICATION OF AN ERD STRONGER THAN 40% (STANDARD METHOD).

|  | C3 | Cz | C4 |
|---|---|---|---|
| $R_1A_1$ | 28,21 | 33,59 | 26,11 |
| $R_1A_2$ | 43,86 | 52,30 | 52,53 |
| $R_1A_3$ | 21,82 | 26,38 | 19,4 |
| $R_2A_1$ | 14,38 | 13,97 | 19,58 |
| $R_2A_2$ | 41,12 | 40,03 | 40,94 |
| $R_2A_3$ | 13,14 | 13,97 | 17,12 |

TABLE III. MEAN AND STANDARD DEVIATION VALUES OF ERD IN THE TRIALS WHERE IT WAS IDENTIFIED (STANDARD METHOD).

|  | C3 |  | Cz |  | C4 |  |
|---|---|---|---|---|---|---|
|  | *mean* | *std* | *mean* | *std* | *mean* | *std* |
| $R_1A_1$ | -65,53 | 14,89 | -64,46 | 15,19 | -64,58 | 14,95 |
| $R_1A_2$ | -69,75 | 15,21 | -67,30 | 15,42 | -67,74 | 15,13 |
| $R_1A_3$ | -66,79 | 15,81 | -66,25 | 16,18 | -66,10 | 15,36 |
| $R_2A_1$ | -68,11 | 16,44 | -66,13 | 15,74 | -66,34 | 16,04 |
| $R_2A_2$ | -69,99 | 15,53 | -69,69 | 15,95 | -71,99 | 15,38 |
| $R_2A_3$ | -67,49 | 15,37 | -66,87 | 14,63 | -67,56 | 16,70 |

It has to be observed here that the ERD identification based on the reference period $R_1$ usually led to higher detection performance, since the baseline energy was evaluated from a longer period of acquisition before the experiment beginning in which the relaxation (and the consequent SMR enhancement) of the subject was thus ensured. However, sometimes the subject conditions at the current trial were so different from those at the beginning that a nearer (in time) reference would be a better choice, in order to avoid many false positive errors. In these cases, the inter-trials periods $R_2$ should be used. The same option would result beneficial in case of large and long-lasting artifacts corrupting the rest, as for example the electrode-polarization ones illustrated before. By their side, pre-trigger times last 500 ms only, then they could show a non-complete relaxation of the patient leading to a number of false negative errors in the identification process. Therefore, both the solutions present their advantages as well as their drawbacks: a trade-off has to be accepted then. As far as in author's procedure the second solution was preferred, the most suitable comparison between the two methods could be accomplished by looking at the fourth row in each of the previous tables and the first two rows of the following ones. Moreover, for what the frequency bands of the second method regard, the bands within the range (9,14) Hz will be reported. Indeed, this corresponds to the range of the individual frequencies for all C3, Cz and C4 for the ($R_2$, $A_1$) pair of intervals.

TABLE IV. PERCENTAGE OF IDENTIFICATION OF AN ERD STRONGER THAN 40% IN THE BAND (11.5, 13.5) HZ (NOVEL METHOD).

|  | left | inter | right |
|---|---|---|---|
| -1.5 s | 10.04 | 40.32 | 10.83 |
| -1 s | 11.77 | 52.14 | 13.34 |
| -0.5 s | 15.95 | 64.44 | 19.35 |
| onset | 19.14 | 73.48 | 24.48 |

TABLE V.  MEAN AND STANDARD DEVIATION VALUES OF ERD IN THE TRIALS WHERE IT WAS IDENTIFIED (NOVEL METHOD).

|        | left   |       | inter  |       | right  |       |
|--------|--------|-------|--------|-------|--------|-------|
|        | mean   | std   | mean   | std   | mean   | std   |
| -1.5 s | -54.88 | 11.05 | -67.18 | 15.65 | -58.04 | 11.69 |
| -1 s   | -52.39 | 9.54  | -59.92 | 12.95 | -50.05 | 8.40  |
| -0.5 s | -49.64 | 7.45  | -59.30 | 12.50 | -51.18 | 8.58  |
| onset  | -48.92 | 7.70  | -61.00 | 12.69 | -51.87 | 9.00  |

## V. DISCUSSION

The comparison established between the outcomes from the two methods will be explained in this section and it is based on the correspondence between the ERD values found in the case $R_2A_1$ in the Pfurtscheller's method and those obtained from the first two intervals of each trial in the second procedure: in both the cases the ERD is estimated as early as the feedback computation can be completed before the actual movement. One of the main differences between the two procedures relies on the identification of the individual frequency: although a semi-qualitative solution was proposed by Pfurtscheller and colleagues, this is needed in the direction of the most suitable customization of the experiment on the particular subject. Moreover, the first method required to choose the most relevant reference and active periods a-priori, while the author's procedure computed the ratio between subsequent intervals allowing to identify the energy decrease in a more precise way during the whole trial course before the actual movement. Finally, filtering with a 2 Hz bandwidth causes about 500 ms delay in the output signal. This could compromise the online operations of the first method as far as the single traces of 1 second were filtered. Despite of these considerations, this method is widely used in the BCI community for the ERD/ERS computation, so it is possible to compare results from several protocols and dataset. On the other side, author's method has the advantage that, thanks to a non optimized inter-trials interval - actually too short to make the subject completely relax - it can be said to be suitable for a more realistic situation of movement: indeed, as in the real world outside the laboratory, the patient cannot wait the time needed to regain a total rest situation after every action. In such a context, the algorithm designed and implemented by authors resulted quite effective whereas the other one would need a longer and an offline procedure to estimate the ERD. For what the spatial distribution of the ERD regards, the first method accounted channels C3, Cz and C4 a priori as the most significant locations to analyze the ERD behavior. This is supported by literature, but it could not hold for every patient. The author's method instead perform an overall analysis on all the signals and computed an average between group of signals at the end of the procedure only, letting the researchers to analyze the cerebral activity in the whole scalp. On expectations' contrary, no lateralization of the activity in the contralateral hemisphere was clearly observed and the desynchronization appeared to be focused over the central part of the scalp where higher values of ERD identification were found. This could be explained because intense communications between the hemispheres could be expected during the accomplishment of a motor task. Then, author's method showed slightly better results with a 70% of identification among the inter-hemispheres signals (second column of Table 4), while only a 40% could be achieved by means of the Pfurtscheller's method in all the channels considered. Moreover, a slight difference between the performance of the left and the right sides of the scalp could be highlighted by the author's method: as expected, the patient moving the left arm shows a slightly more pronounced ERD activity over the right sensorimotor part of the scalp with a consequent higher level of identifications. To conclude, several considerations could be made to complete the discussion: first of all, it is generally difficult to assess the spatial distribution of the ERD by means of the EEG because of the volume conduction effects inside the brain; to this purpose, a lower threshold (for example, 20%) could support the decoding of the ERD and could be considered for further analysis. Finally, other data from other stroke patients and from healthy subjects are currently being analyzed to confirm these preliminary findings.

REFERENCES

[1] J.J. Vidal, "Toward Direct Brain-Computer Communications", *Annu. Rev. Biophys. Bioeng.*, vol. 2, pp. 157–180, 1973.
[2] L.A. Farwell and E. Donchin, "Talking off the top of your head: toward a mental prosthesis utilizing event-related brain potentials, *EEG Clin. Neurophysiol.*, vol. 70, n. 6, 510-523, 1988.
[3] N. Birbaumer, N. Ghanayim, T. Hinterberger, I. Iversen, B. Kotchoubey, A. Kubler, J. Perelmouter, E. Taub and H. Flor, "A spelling device for the paralysed", *Nature*, 1999.
[4] L.R. Hochberg, D. Bacher, B. Jarosiewicz, N.Y. Masse, J.D. Simeral, J. Vogel, S. Haddadin, J. Liu, S.S. Cash, P. Van Der Smagt and J.P. Donoghue, "Reach and grasp by people with tetraplegia using a neurally controlled robotic arm", *Nature*, vol. 485, n. 7398, pp. 372-375, 2012.
[5] J.L. Collinger, S. Foldes, T.M. Bruns, B. Wodlinger, R. Gaunt and D.J. Weber, "Neuroprosthetic technology for individuals with spinal cord injury", *J. Spinal Cord. Med.*, vol. 36, n. 4, pp. 258-272, 2013.
[6] G. Pfurtscheller and A. Aranibar, "Event-related cortical desynchronization detected by power measurements of scalp EEG", *EEG Clin. Neurophysiol.*, vol. 42, n. 6, pp. 817-826, 1977.
[7] J.R. Wolpaw, N. Birbaumer, D.J. McFarland, G. Pfurtscheller and T.M. Vaughan, "Brain-Computer interfaces for communication and control", *Clin. Neurophysiol.*, vol. 113, pp. 767-791, 2002.
[8] E.E. Fetz, "Operant Conditioning of Cortical Unit Activity", *Science*, vol. 163, pp. 955-958, 1969.
[9] G. Pfurtscheller and F.H. Lopes Da Silva, "Event-related EEG/MEG synchronization and desynchronization: basic principles", *Clin. Neurophysiol.*, vol. 110, n. 11, pp. 1842-1857, 1999.
[10] C. Neuper, M. Wortz and G. Pfurtscheller, "ERD/ERS patterns reflecting sensorimotor activation and deactivation", *Prog. Brain Res.*, vol. 159, pp. 211-222, 2006.
[11] G. Cisotto, S. Silvoni, M. Cavinato, F. Piccione and S. Pupolin, "Brain computer interface in chronic stroke: an application of sensorimotor closed-loop and contingent force feedback", *Proc. IEEE ICC'13*, pp. 2972-2976, 2013.